\documentclass[aps,prd,twocolumn,superscriptaddress,showpacs]{revtex4-1}
\usepackage{graphicx}
\usepackage{amsmath}
\usepackage{array}
\usepackage{multirow}
\usepackage{comment}
\usepackage{lineno}
\usepackage[usenames, dvipsnames]{color}

\begin{document}

\preprint{}

\title{Measurement of the longitudinal spin asymmetries for weak boson production in proton--proton collisions at $\sqrt{s}$ = 510\,GeV}


\affiliation{Abilene Christian University, Abilene, Texas   79699}
\affiliation{AGH University of Science and Technology, FPACS, Cracow 30-059, Poland}
\affiliation{Alikhanov Institute for Theoretical and Experimental Physics, Moscow 117218, Russia}
\affiliation{Argonne National Laboratory, Argonne, Illinois 60439}
\affiliation{Brookhaven National Laboratory, Upton, New York 11973}
\affiliation{University of California, Berkeley, California 94720}
\affiliation{University of California, Davis, California 95616}
\affiliation{University of California, Los Angeles, California 90095}
\affiliation{University of California, Riverside, California 92521}
\affiliation{Central China Normal University, Wuhan, Hubei 430079 }
\affiliation{University of Illinois at Chicago, Chicago, Illinois 60607}
\affiliation{Creighton University, Omaha, Nebraska 68178}
\affiliation{Czech Technical University in Prague, FNSPE, Prague 115 19, Czech Republic}
\affiliation{Technische Universit\"at Darmstadt, Darmstadt 64289, Germany}
\affiliation{E\"otv\"os Lor\'and University, Budapest, Hungary H-1117}
\affiliation{Frankfurt Institute for Advanced Studies FIAS, Frankfurt 60438, Germany}
\affiliation{Fudan University, Shanghai, 200433 }
\affiliation{University of Heidelberg, Heidelberg 69120, Germany }
\affiliation{University of Houston, Houston, Texas 77204}
\affiliation{Huzhou University, Huzhou, Zhejiang 313000}
\affiliation{Indiana University, Bloomington, Indiana 47408}
\affiliation{Institute of Physics, Bhubaneswar 751005, India}
\affiliation{University of Jammu, Jammu 180001, India}
\affiliation{Joint Institute for Nuclear Research, Dubna 141 980, Russia}
\affiliation{Kent State University, Kent, Ohio 44242}
\affiliation{University of Kentucky, Lexington, Kentucky 40506-0055}
\affiliation{Lawrence Berkeley National Laboratory, Berkeley, California 94720}
\affiliation{Lehigh University, Bethlehem, Pennsylvania 18015}
\affiliation{Max-Planck-Institut f\"ur Physik, Munich 80805, Germany}
\affiliation{Michigan State University, East Lansing, Michigan 48824}
\affiliation{National Research Nuclear University MEPhI, Moscow 115409, Russia}
\affiliation{National Institute of Science Education and Research, HBNI, Jatni 752050, India}
\affiliation{National Cheng Kung University, Tainan 70101 }
\affiliation{Nuclear Physics Institute of the CAS, Rez 250 68, Czech Republic}
\affiliation{Ohio State University, Columbus, Ohio 43210}
\affiliation{Institute of Nuclear Physics PAN, Cracow 31-342, Poland}
\affiliation{Panjab University, Chandigarh 160014, India}
\affiliation{Pennsylvania State University, University Park, Pennsylvania 16802}
\affiliation{National Research Centre ``Kurchatov Institute" -- Institute of High Energy Physics, Protvino 142281, Russia}
\affiliation{Purdue University, West Lafayette, Indiana 47907}
\affiliation{Pusan National University, Pusan 46241, Korea}
\affiliation{Rice University, Houston, Texas 77251}
\affiliation{Rutgers University, Piscataway, New Jersey 08854}
\affiliation{Universidade de S\~ao Paulo, S\~ao Paulo, Brazil 05314-970}
\affiliation{University of Science and Technology of China, Hefei, Anhui 230026}
\affiliation{Shandong University, Qingdao, Shandong 266237}
\affiliation{Shanghai Institute of Applied Physics, Chinese Academy of Sciences, Shanghai 201800}
\affiliation{Southern Connecticut State University, New Haven, Connecticut 06515}
\affiliation{State University of New York, Stony Brook, New York 11794}
\affiliation{Temple University, Philadelphia, Pennsylvania 19122}
\affiliation{Texas A\&M University, College Station, Texas 77843}
\affiliation{University of Texas, Austin, Texas 78712}
\affiliation{Tsinghua University, Beijing 100084}
\affiliation{University of Tsukuba, Tsukuba, Ibaraki 305-8571, Japan}
\affiliation{United States Naval Academy, Annapolis, Maryland 21402}
\affiliation{Valparaiso University, Valparaiso, Indiana 46383}
\affiliation{Variable Energy Cyclotron Centre, Kolkata 700064, India}
\affiliation{Warsaw University of Technology, Warsaw 00-661, Poland}
\affiliation{Wayne State University, Detroit, Michigan 48201}
\affiliation{Yale University, New Haven, Connecticut 06520}

\author{J.~Adam}\affiliation{Creighton University, Omaha, Nebraska 68178}
\author{L.~Adamczyk}\affiliation{AGH University of Science and Technology, FPACS, Cracow 30-059, Poland}
\author{J.~R.~Adams}\affiliation{Ohio State University, Columbus, Ohio 43210}
\author{J.~K.~Adkins}\affiliation{University of Kentucky, Lexington, Kentucky 40506-0055}
\author{G.~Agakishiev}\affiliation{Joint Institute for Nuclear Research, Dubna 141 980, Russia}
\author{M.~M.~Aggarwal}\affiliation{Panjab University, Chandigarh 160014, India}
\author{Z.~Ahammed}\affiliation{Variable Energy Cyclotron Centre, Kolkata 700064, India}
\author{I.~Alekseev}\affiliation{Alikhanov Institute for Theoretical and Experimental Physics, Moscow 117218, Russia}\affiliation{National Research Nuclear University MEPhI, Moscow 115409, Russia}
\author{D.~M.~Anderson}\affiliation{Texas A\&M University, College Station, Texas 77843}
\author{R.~Aoyama}\affiliation{University of Tsukuba, Tsukuba, Ibaraki 305-8571, Japan}
% ...

\collaboration{STAR Collaboration}\noaffiliation



\date{\today}

\begin{abstract}
We report new STAR measurements of the single-spin asymmetries $A_L$ for $W^+$ and $W^-$ bosons produced in polarized proton--proton collisions at $\sqrt{s}$ = 510\,GeV as a function of the decay-positron and decay-electron pseudorapidity. The data were obtained in 2013 and correspond to an integrated luminosity of 250 pb$^{-1}$. The results are combined with previous results obtained with 86 pb$^{-1}$. A comparison with theoretical expectations based on polarized lepton-nucleon deep-inelastic scattering and prior polarized proton--proton data suggests a difference between the $\bar{u}$ and $\bar{d}$ quark helicity distributions for $0.05 < x < 0.25$.  In addition, we report new results for the double-spin asymmetries $A_{LL}$ for $W^\pm$\!, as well as $A_L$ for $Z/\gamma^*$ production and subsequent decay into electron--positron pairs.
\end{abstract}

\pacs{13.38.Be, 13.38.Dg, 13.88.+e, 14.20.Dh, 24.85.+p}

\maketitle


Understanding the spin structure of the proton in terms of its quark, antiquark, and gluon constituents is of fundamental interest. This description is commonly done using polarized parton distribution functions (PDFs), which can be determined using perturbative QCD techniques and global analyses of data from polarized deep-inelastic lepton-nucleon scattering (DIS) experiments and from high-energy polarized proton--proton scattering experiments at the Relativistic Heavy-Ion Collider (RHIC).
Recent examples of such PDFs are given in Refs.~\cite{Nocera:2014gqa,deFlorian:2014yva}.
The data from leptonic $W$-decays in polarized proton--proton collisions at RHIC~\cite{Aggarwal:2010vc,Adamczyk:2014xyw, Adare:2010xa, Adare:2015gsd, Adare:2018csm} provide constraints in these global analyses, which now
show a flavor asymmetry in the light sea-quark polarizations for parton momentum fractions, $0.05 < x < 0.25,$ at hard perturbative scales.
The existence of such an asymmetry in the polarized PDFs has been searched for directly in semi-inclusive DIS experiments~\cite{Adeva:1998qz, Airapetian:2004zf, Alekseev:2010ub} but had thus far been established only in the case of the unpolarized PDFs.
There, Drell-Yan measurements~\cite{Baldit:1994jk,Towell:2001nh} and DIS measurements~\cite{Arneodo:1996qe, Ackerstaff:1998sr}, in particular, have reported  large enhancements in the ratio of $\bar{d}$ over $\bar{u}$ antiquark distributions.
This has provided a strong impetus for theoretical modeling~\cite{Kumano:1997cy} and renewed measurement~\cite{Reimer:2011zza}.
Considerable progress is being made also in lattice-QCD~\cite{Lin:2017snn}.

The leptonic $W^+\rightarrow e^+ \nu$ and $W^-\rightarrow e^- \bar{\nu}$ decay channels provide sensitivity to the helicity distributions of the quarks, $\Delta u$ and $\Delta d$, and antiquarks, $\Delta\bar{u}$ and $\Delta\bar{d}$, that is free of uncertainties associated with non-perturbative fragmentation.
The cross-sections are well described~\cite{STAR:2011aa}.
The primary observable is the longitudinal single-spin asymmetry $A_L \equiv (\sigma_+-\sigma_- )/ (\sigma_++\sigma_-)$ where $\sigma_{+(-)}$ is the cross-section when the helicity of the polarized proton beam is positive (negative).
At leading order,
\begin{equation} 
A_L^{W^+}(y_W) \propto \frac{ \Delta \bar{d}(x_1) u(x_2) - \Delta u(x_1) \bar{d}(x_2) }{\bar{d}(x_1) u(x_2) + u(x_1) \bar{d}(x_2)},
\label{Eqn:ALWp}
\end{equation}
\begin{equation}
A_L^{W^-}(y_W) \propto \frac{\Delta \bar{u}(x_1) d(x_2)-\Delta d(x_1) \bar{u}(x_2)}{\bar{u}(x_1)d(x_2) + d(x_1) \bar{u}(x_2)},
\label{Eqn:ALWn}
\end{equation} 
where $x_1~(x_2)$ is the momentum fraction carried by the colliding quark or antiquark in the polarized (unpolarized) beam.
$A_L^{W^+}$ ($A_L^{W^-}$) approaches $-\Delta u/u$ ($-\Delta d/d$) in the very forward region of $W$ rapidity, $y_W \gg 0$, and $\Delta \bar{d}/\bar{d}$ ($\Delta \bar{u}/\bar{u}$) in the very backward region of $W$ rapidity, $y_W \ll 0$.
The observed positron and electron pseudorapidities, $\eta_e$, are related to $y_W$ and to the decay angle of the positron and electron in the $W$ rest frame~\cite{Bunce:2000uv}.
Higher-order corrections to $A_L(\eta_e)$ are known~\cite{Ringer:2015oaa,Nadolsky2003,deFlorian:2010aa} and have been incorporated into the aforementioned global analyses.

In this Rapid Communication, we report new measurements of the single-spin asymmetries for decay positrons and electrons from $W^\pm$ bosons produced in longitudinally-polarized proton--proton collisions at a center-of-mass energy of $\sqrt{s}$ = 510\,GeV.
In addition, we report new results for the double-spin asymmetries $A_{LL}$ for $W^\pm$ and $A_L$ for $Z/\gamma^*$ production.
The data were recorded in the year 2013 by the STAR collaboration and correspond to an integrated luminosity of about 250 pb$^{-1}$.
The polarizations of the two incident proton beams were measured using Coulomb-nuclear interference proton--carbon polarimeters, which were calibrated with a polarized hydrogen gas-jet target~\cite{RHICpol}. The luminosity-weighted beam polarization was $P = 0.56$, with a relative scale uncertainty of 3.3\% for the single-beam polarization and 6.4\% for the product of the polarizations from both beams.
The figure-of-merit, $P^2\mathcal{L}$ for single-spin asymmetry measurements, is higher by a factor of three for the 2013 data compared to the results~\cite{Adamczyk:2014xyw} from the 2011 and 2012 data.

This measurement and analysis made use of essentially the same apparatus and techniques as described in Refs.~\cite{Aggarwal:2010vc,STAR:2011aa,Adamczyk:2014xyw}.
As before, the subsystems of the STAR detector~\cite{NIM} used in this measurement are the Time Projection Chamber~\cite{TPC} (TPC), which provides charged particle tracking for pseudorapidities $|\eta| \lesssim 1.3$, and the Barrel~\cite{BEMC} and Endcap~\cite{EEMC} Electromagnetic Calorimeters (BEMC, EEMC). These lead-scintillator sampling calorimeters are segmented into optically isolated towers that cover the full azimuthal angle, $\phi$, for mid and forward pseudorapidity, $|\eta|<1.0$ and $1.1<\eta < 2.0$, respectively.  They provide the online triggering requirements to initiate the data recording.
The trigger accepted events 
if  a transverse energy $E_T > 12\ (10)\, \mathrm{GeV}$ was observed in a region $\Delta\eta\times\Delta\phi \simeq 0.1 \times 0.1$ of the BEMC (EEMC).
Events were kept in the analysis if their collision vertex along the beam axis, determined from tracks reconstructed in the TPC, was within $\pm\,100$\,cm of the center of the STAR detector. The vertex distribution along the beam axis was approximately Gaussian with an RMS width of 47\,cm.

The $W^\pm$ bosons were detected via their decay into positrons and electrons, $W^+\rightarrow e^+ \nu$ and $W^-\rightarrow e^- \bar{\nu}$.
These events are characterized by an isolated $e^+$ or $e^-$ with high transverse momentum, $p_T$, accompanied by a high $p_T$ neutrino, $\nu$, or antineutrino, $\bar{\nu}$.  Since the $\nu$ and $\bar{\nu}$ escape detection, this leads to a characteristically large $p_T$ imbalance in these events.

Candidate $W$-decay positrons or electrons were identified at mid-rapidity (forward rapidity) by a high $p_T$ TPC track associated with the primary event collision vertex pointing to a matching tower cluster in the BEMC (EEMC) with high energy.  Candidate tracks at mid-rapidity (forward rapidity) were required to have at least 15\,(5) TPC hits to ensure good track quality, and the ratio of the number of hits in the fit to the number of possible hits was required to be more than 0.51 to avoid splitting tracks.  
A threshold was imposed on the transverse momentum of the particle track, $p_T > 10\,(7)\,\mathrm{GeV/}c$.

Of the four possible 2 $\times$ 2 calorimeter tower clusters containing the tower that was hit at its front face by the high-$p_T$ positron or electron, the cluster with the largest total energy was used to determine the positron or electron transverse energy, $E^e_T$.  This energy was required to exceed 14 GeV.   The distance between the track and the center position of the tower cluster was required to be less than 7\,(10)\,cm at the front face of the BEMC (EEMC).

Unlike background events, signal events have a characteristic isolated transverse energy deposit from the decay positron or electron of about 40\,GeV, approximately half the $W$ mass, and a large imbalance in the total observed transverse energy as mentioned above. QCD backgrounds were suppressed using selections based on kinematic and topological differences between leptonic $W$-decay events and QCD processes. To identify isolated high-$p_T$ decay positrons or electrons, and discriminate against jets, the ratio of $E^e_T$ to the total energy in a $4\times4$ BEMC (EEMC) cluster centered on and including the candidate $2\times2$ tower cluster was required to be greater than $95\,(96)\%$. In addition, the ratio of $E^e_T$ to the transverse energy $E^{\Delta R < 0.7}_T$ in a cone of radius of $\Delta R = \sqrt{\Delta\eta^2 + \Delta\phi^2} <  0.7$ around the candidate track was required to be greater than
88\%.
The transverse energy $E^{\Delta R < 0.7}_T$ was determined by summing the BEMC and EEMC $E_T$ and the TPC track $p_T$ within the cone.  This selection thus suppressed jet-like events.
In addition, in the EEMC acceptance, an isolation cut based on the energy deposited in the two layers of the EEMC Shower Maximum Detector (ESMD)~\cite{EEMC} was used.  The ESMD can be used to measure the transverse profile of the electromagnetic shower and thereby discriminate between the narrow transverse profile of an isolated (signal) positron or electron shower and the typically wider distribution observed in QCD (background) events. This was done by requiring that the ratio of total energy deposited in ESMD strips within $\pm$1.5\,cm of the central strip pointed to by a TPC track to the energy deposited in strips within $\pm$10\,cm, $R_\mathrm{ESMD}$, was greater than 0.7.

In addition, the characteristic transverse energy imbalance of signal events was used to further suppress backgrounds.  A $p_T$-balance vector, $\vec{p}_T^{\ \mathrm{bal}}$, defined as the vector-sum of the decay positron or electron candidate $\vec{p}^{\ e}_T$ vector plus the sum of the $\vec{p}_T$ vectors for all reconstructed jets whose axes are outside a cone radius of $\Delta R$ = 0.7 around the candidate decay positron or electron, was computed for each event.   Jets were reconstructed for this purpose using an anti-$k_T$ algorithm~\cite{Cacciari:2008gp} with a resolution parameter $R = 0.6$ from towers (tracks) with $E_T$ ($p_T$) $> 0.2$ GeV($/c$). Reconstructed jets were required to have $p_T > 3.5$ GeV$/c$.
A scalar signed $p_T$-balance variable, defined as $(\vec{p}_T^{\ e}\cdot\,\vec{p}^{\ \mathrm{bal}}_T)/|\vec{p}^{\ e}_T|$, was then computed and required to be larger than 14 (20)\,GeV/$c$ for candidate events in the BEMC (EEMC) to be retained in the analysis.
Complementary to the signed $p_T$-balance cut, 
it was required that the total transverse energy opposite in azimuth to the candidate positron or electron in the BEMC, $-0.7 <  \Delta\phi -\pi < 0.7,$ did not exceed 11\,GeV. This further reduced QCD dijet background in cases when a sizable fraction of the energy for one of the jets was not observed due to detector effects. 

\begin{figure}[ht]
\centering
\includegraphics[width=8.6cm]{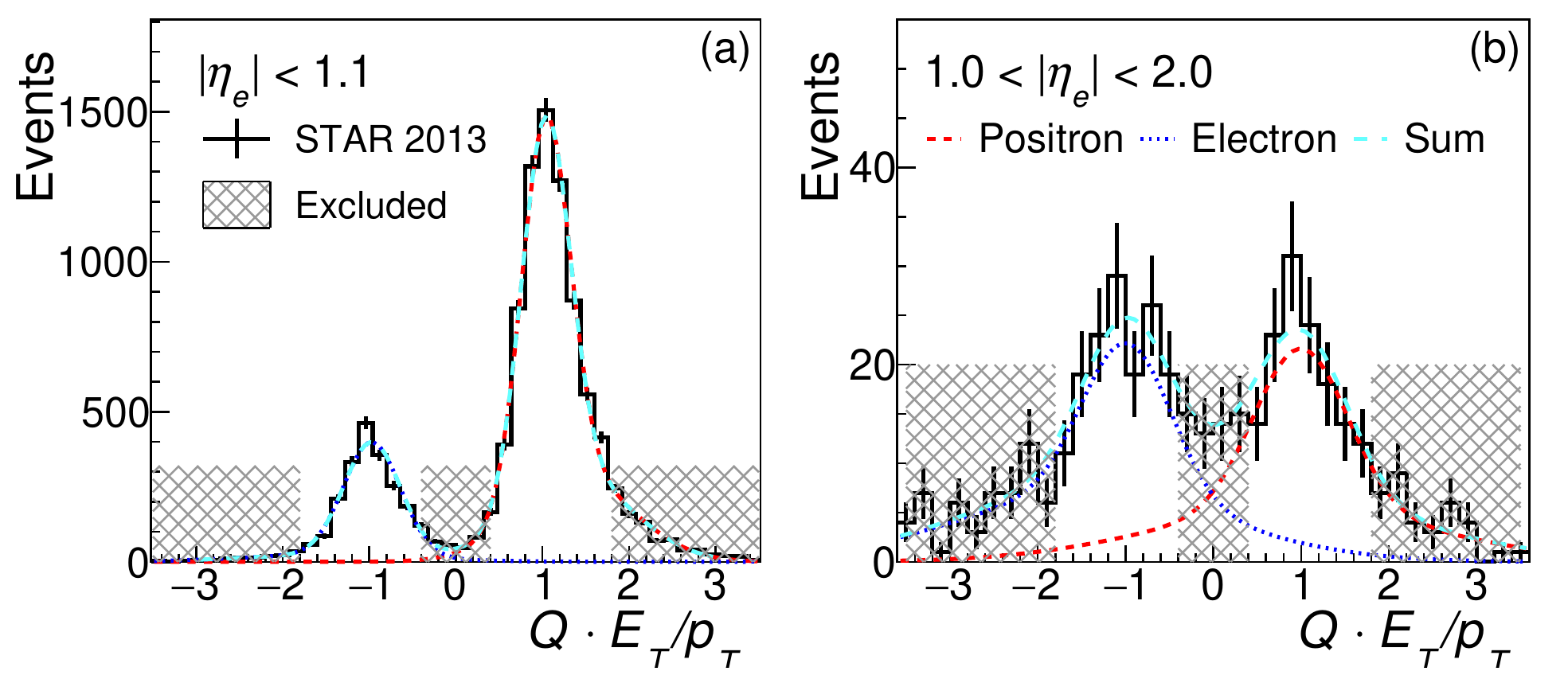}
\caption{\label{fig:charge} Distribution of the product of $Q$, the TPC reconstructed charge-sign, and $E_T/p_T$ in the BEMC (a) and EEMC (b) regions.
The positron (red) and electron (blue)  candidate events have been fitted with double-Gaussian distributions. The excluded regions are marked by hatched shades.}
\end{figure}

Candidate positrons or electrons that passed the above selection cuts were then sorted by charge-sign, determined from the curvature of the TPC tracks in the  solenoidal magnetic field.  Figure~\ref{fig:charge}a (b) shows the distribution of the reconstructed charge-sign, $Q = \pm 1$, multiplied by the ratio of $E^e_T$ observed in the BEMC (EEMC) to $p^e_T$ determined with the TPC for events in the signal region, $25 <  E_T^e <  50\,\mathrm{GeV}$. The relative yields of the $W^+$ and $W^-$ follow the pseudorapidity dependence of the cross-section ratio.  The distributions were each fitted with two double-Gaussian template shapes, determined from a Monte Carlo simulated $W$ sample, to estimate the reconstructed charge-sign purity.
The amplitudes of the Gaussians were fitted to the data, as was the central position of the narrower Gaussian in each of the templates.  The remaining parameters were fixed by studies in which simulated $W^+\rightarrow e^+ \nu$ and $W^-\rightarrow e^- \bar{\nu}$ events were embedded (c.f. the paragraph below) in zero-bias data.
The hatched regions, $|Q\cdot E_T/p_T| < 0.4$ and $|Q\cdot E_T/p_T| > 1.8$, were excluded 
to remove tracks with poorly reconstructed $p_T$ and to reduce contamination from events with opposite charge-sign.  This contamination is negligible at mid-rapidity, but increases to  9.6\% and 12.0\% for $W^+$ and $W^-$ candidate events, respectively, in the EEMC region.
The forward $A_L$ values were corrected for this contamination using the  asymmetries observed in the data.

\begin{widetext}

\begin{figure}[h]
\centering
\includegraphics[width=17.8cm]{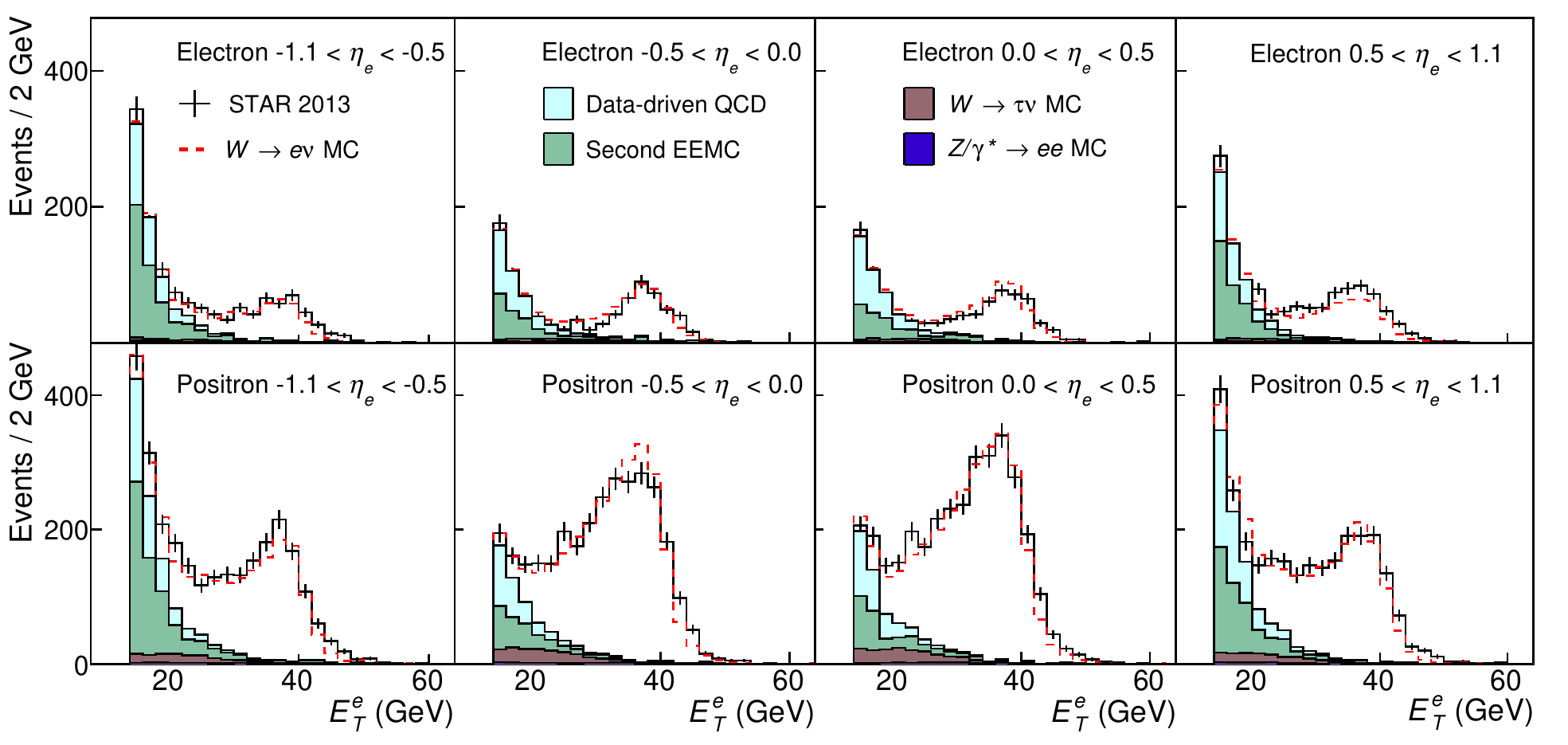}
\caption{\label{fig:bemc} $E_T^e$ distribution of  electron (top) and positron (bottom) candidates (black crosses), background contributions, and sum of backgrounds and $W \rightarrow e\nu$ MC signal (red-dashed) in the BEMC region.}
\end{figure}

\end{widetext}

Figure~\ref{fig:bemc} shows the distributions of $W^+$ and $W^-$ yields as a function of $E^e_T$ for the four central $\eta_e$ intervals considered in this analysis, along with the estimated residual background contributions from electroweak and QCD processes. 
The residual electroweak backgrounds are predominantly due to $W^\pm\rightarrow\tau^\pm\nu_\tau$ and $Z/{\gamma^*}\rightarrow e^+ e^-$. These contributions were estimated from Monte Carlo simulations, using events generated with \textsc{pythia 6.4.28}~\cite{Sjostrand:2006za} and the ``Perugia 0" tune~\cite{Skands:2010ak} that passed through a \textsc{geant 3}~\cite{Brun:1994aa} model of the STAR detector, and were subsequently embedded into STAR zero-bias data.  The simulated samples were normalized to the $W$ data using the known integrated luminosity.  The \textsc{tauola} package was used for the polarized $\tau^\pm$ decay~\cite{Golonka:2003xt}.
Residual QCD dijet background in which one of the jets pointed to  uninstrumented pseudorapidity regions was estimated using two separate procedures.
The contribution from $e^\pm$ candidate events with an opposite-side jet fragment in the uninstrumented region $-2 < \eta < -1.1$ was estimated by studying such data in the EEMC, which instruments the region $1.1 < \eta < 2$. 
This is referred to as the ``Second EEMC" procedure.
Residual background from the uninstrumented region $|\eta|>2$ was estimated by studying events that satisfy all isolation criteria, but do not satisfy the cuts on the scalar signed $p_T$-balance variable. This is referred to as the ``Data-driven QCD" procedure. 
 To assess the background remaining in the signal region, the $E_T$ distribution of this background-dominated sample was normalized to the signal candidate distribution  that remained after all other background contributions had been removed for $E_T$ values between 14\,GeV and 18\,GeV.  Additional aspects of both procedures are described in  Refs~\cite{Aggarwal:2010vc,STAR:2011aa}.

\begin{figure}[tb]
\centering
\includegraphics[width=8.6cm]{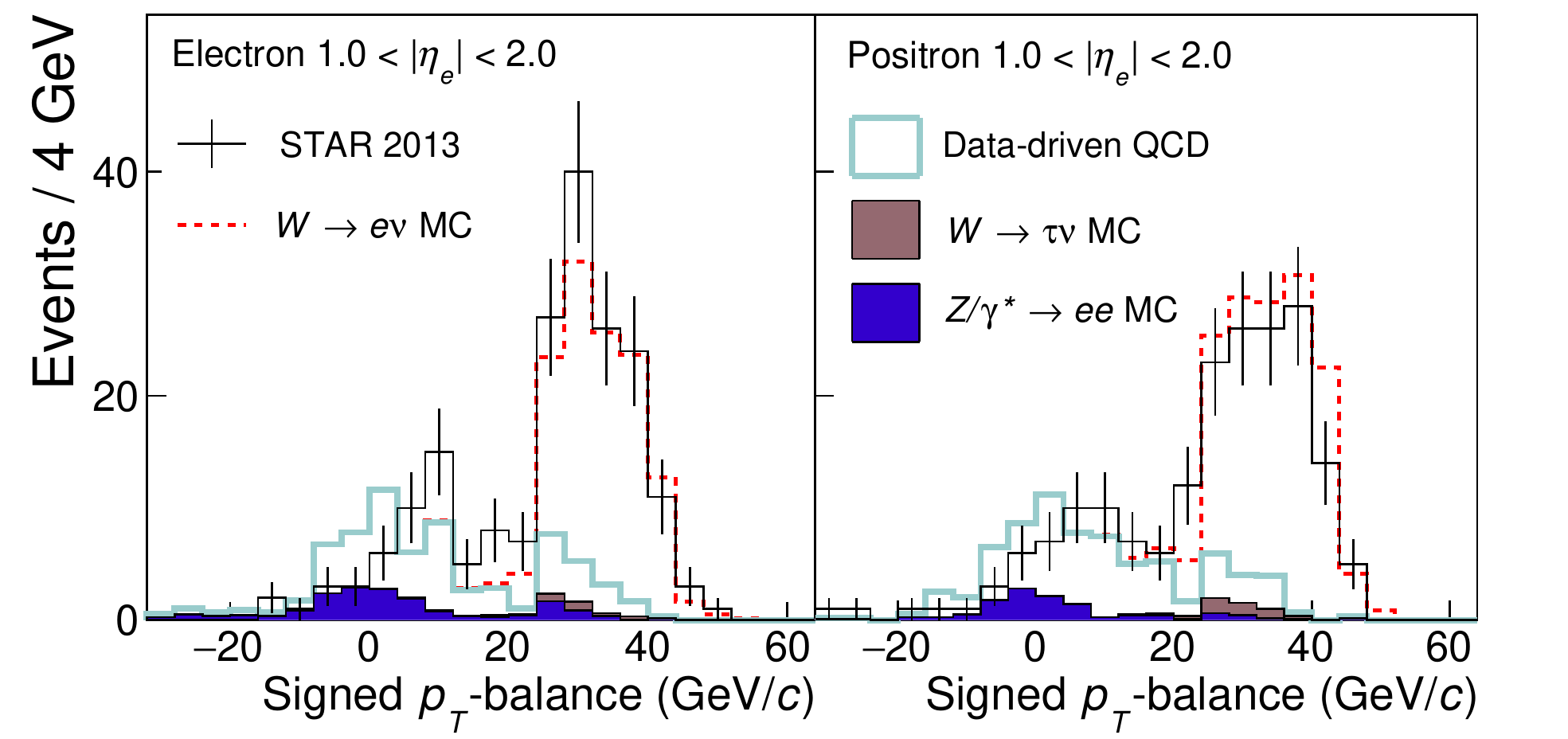}
\caption{\label{fig:eemc} Signed $p_{T}$-balance distribution for electron (left) and positron (right) candidates (black crosses), background contributions, and sum of backgrounds and $W \rightarrow e\nu$ MC signal (red-dashed) in the EEMC region.}
\end{figure}

Figure~\ref{fig:eemc} shows the charge-separated distributions in the EEMC region as a function of the signed $p_{T}$-balance variable, together with the estimated residual background contributions.
Residual electroweak backgrounds for these regions were estimated in the same way as for the mid-rapidity data.  Residual QCD backgrounds were estimated using the ESMD, where the isolation parameter $R_\mathrm{ESMD}$  was required to be less than 0.6 for  QCD background events.
The shape was determined for each charge-sign separately and normalized to the measured yield in the region where the signed $p_T$-balance variable was between -8 and 8\,GeV/$c$.  This region is dominated by QCD backgrounds.

At RHIC, there are four helicity configurations for the two longitudinally-polarized proton beams: $++$, $+-$, $-+$, and $--$.
The data from these four configurations can be combined such that the net polarization for one beam effectively averages to zero, while maintaining high polarization in the other.
The longitudinal single-spin asymmetry $A_L$ for the combination in which the first beam is polarized and the second carries no net polarization was determined from:
\begin{equation}
A_L = \frac{1}{\beta P}\frac{R_{++}N_{++}+R_{+-}N_{+-}-R_{-+}N_{-+}-R_{--}N_{--}}{R_{++}N_{++}+R_{+-}N_{+-}+R_{-+}N_{-+}+R_{--}N_{--}}
\end{equation}
where $\beta$ is the signal purity, $P$ is the average beam polarization, and $R$ and $N$ are the normalizations for relative luminosity and the raw $W^\pm$ yields, respectively, for the helicity configurations indicated by the subscripts. The relative luminosities were obtained from a large QCD sample that exhibits no significant single-spin asymmetry. Typical values were between 0.993 and 1.009. The purity was evaluated from the aforementioned signal and background contributions and was found to be between 83\% and 98\%.
$A_L$ was determined in a similar way for the combination in which the second beam is polarized and the first carries no net polarization, and the values for the two combinations were then combined.

\begin{figure}[tb]
\centering
\includegraphics[width=8.6cm]{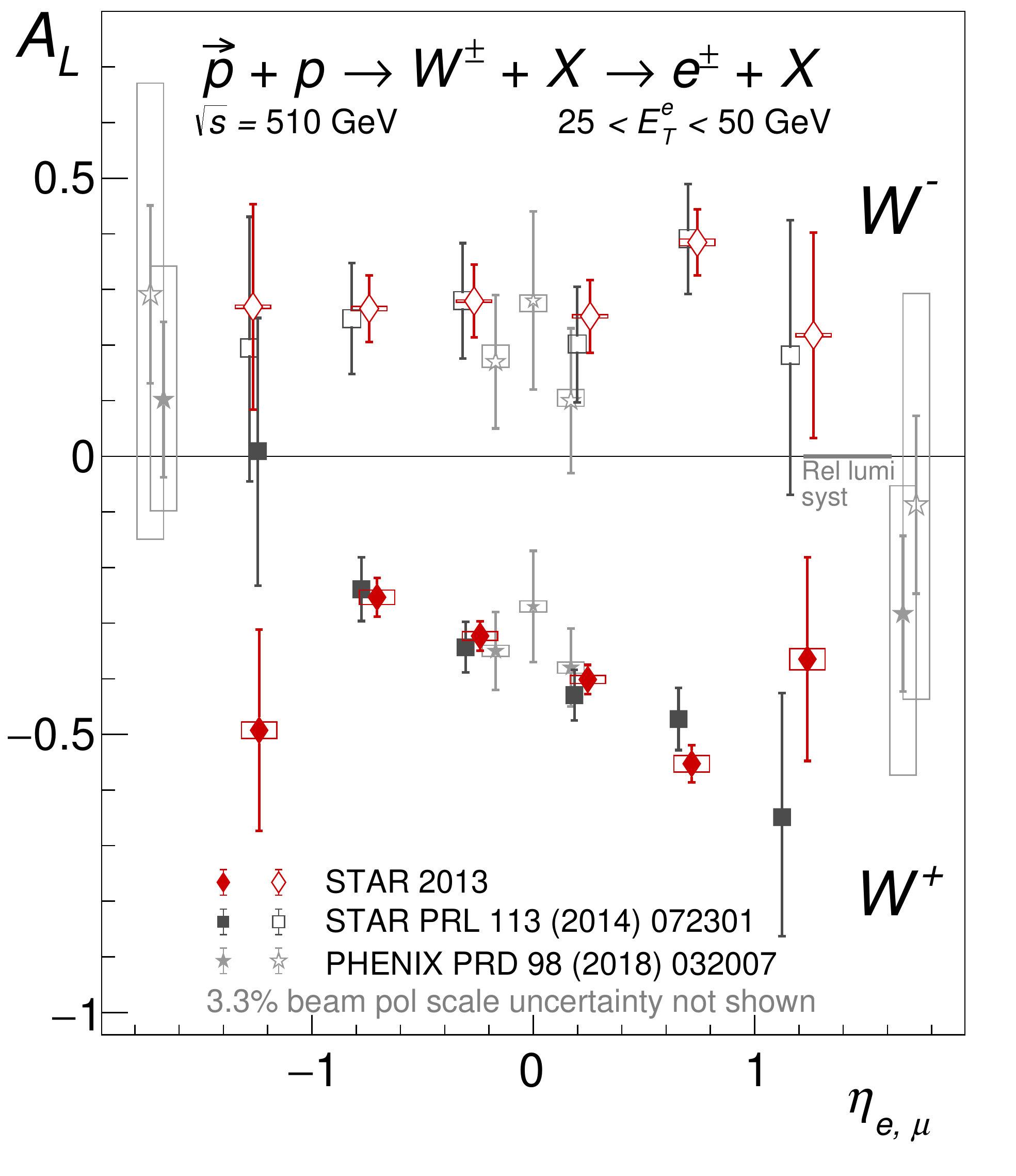}
\caption{\label{fig:money} Longitudinal single-spin asymmetries, $A_L$, for $W^\pm$ production as a function of the positron or electron pseudorapidity, $\eta_e$, separately for the STAR 2011+2012 (black squares) and 2013 (red diamonds) data samples for 25 $< E_T^e <$ 50\,GeV. The 2011+2012 results have been offset to slightly smaller $\eta$ values for clarity.
Shown also are the final asymmetries for high-energy decay leptons from $W$ and $Z/\gamma^*$ production from the PHENIX central arms as a function of $\eta_e$ and from the muon-arms as a function of $\eta_\mu$ with their statistical and systematic uncertainties~\cite{Adare:2015gsd,Adare:2018csm}.
}
\end{figure}

The $A_L$ results for $W^+$ and $W^-$ from the data sample recorded by STAR in 2013 are shown in Fig.~\ref{fig:money} as a function of $\eta_e$. The vertical error bars show the size of the statistical uncertainties, including those associated with the correction for the wrong charge-sign in the case of the points at $|\eta_e| \simeq$ 1.2.  The previously published STAR data~\cite{Adamczyk:2014xyw} are shown for comparison.
Shown also are the $A_L$ data on high-energy forward decay muons and mid-rapidity positrons or electrons from combined $W$ and $Z/\gamma^*$ production by the PHENIX experiment with their statistical and systematic uncertainties as a function of $\eta_\mu$ and $\eta_e$, respectively~\cite{Adare:2015gsd,Adare:2018csm}.

The size of systematic uncertainties associated with BEMC and EEMC gain calibrations (5\% variation) and the data-driven QCD background are indicated by the boxes. The gray band shown along the $A_L$ = 0 line indicates the size of the systematic uncertainty from the determination of relative luminosity, and is correlated among all the points.  The 3.3\% relative systematic uncertainty from beam polarization measurement is not shown.
Table~\ref{tab:WAL2013} gives the results for $A_L$, as well as for the longitudinal double-spin asymmetry $A_{LL} \equiv (\sigma_{++}+\sigma_{--}-\sigma_{+-}-\sigma_{-+})/(\sigma_{++}+\sigma_{--}+\sigma_{+-}+\sigma_{-+})$, where the subscripts denote the helicity configurations.
The new $W^\pm$ $A_{LL}$ data are consistent with previously published STAR data~\cite{Adamczyk:2014xyw} and have better precision.
$W^\pm$ $A_{LL}$ is sensitive to quark and antiquark polarizations, albeit less so than $A_L$, and has been proposed for tests of consistency and positivity constraints~\cite{Chang:2014jba,Kang:2011qz}.

\begin{table*}[tb]
  \begin{center}
    \begin{tabular}{c@{\ \ \ }c@{\ \ \ }c@{\ \ \ }c@{\ \ \ }c@{\ \ \ }c} \hline \hline
    	& \multirow{2}{*}{ $\left<\eta_e\right>$} & \multicolumn{2}{c}{$A_L \pm \sigma_\mathrm{stat} \pm \sigma_\mathrm{syst}$ } & \multicolumn{2}{c}{ $A_{LL}\pm \sigma_\mathrm{stat} \pm \sigma_\mathrm{syst}$} \\ \cline{3-6}
				& & 2013 &  2011--2013  & 2013  &  2011--2013 \\ \hline
  	\multirow{6}{*}{$W^+$}
	&-1.24   &-0.493 $\pm$ 0.181 $\pm$ 0.022 &-0.312 $\pm$ 0.145 $\pm$ 0.017&\\
	&-0.72   &-0.255 $\pm$ 0.035 $\pm$ 0.016 &-0.251 $\pm$ 0.030 $\pm$ 0.014& -- & -- \\
	&-0.25   &-0.327 $\pm$ 0.027 $\pm$ 0.014 &-0.331 $\pm$ 0.023 $\pm$ 0.014&\\
	&\ 0.25  &-0.406 $\pm$ 0.027 $\pm$ 0.016 &-0.412 $\pm$ 0.023 $\pm$ 0.016&\ 0.039 $\pm$ 0.049 $\pm$ 0.014 &\ 0.016 $\pm$ 0.042 $\pm$ 0.011 \\
	&\ 0.72  &-0.557 $\pm$ 0.034 $\pm$ 0.024 &-0.534 $\pm$ 0.029 $\pm$ 0.022&\ 0.049 $\pm$ 0.063 $\pm$ 0.014 &\ 0.072 $\pm$ 0.054 $\pm$ 0.011 \\
	&\ 1.24  &-0.365 $\pm$ 0.183 $\pm$ 0.023 &-0.482 $\pm$ 0.140 $\pm$ 0.022& -0.052 $\pm$ 0.331 $\pm$ 0.044 &\ 0.000 $\pm$ 0.262 $\pm$ 0.028 \\ \hline
	\multirow{6}{*}{$W^-$}
	&-1.27   &\ 0.269 $\pm$ 0.185 $\pm$ 0.010 &\ 0.241 $\pm$ 0.146 $\pm$ 0.010& \\ 
	&-0.74   &\ 0.264 $\pm$ 0.060 $\pm$ 0.010 &\ 0.260 $\pm$ 0.051 $\pm$ 0.010&  -- & -- \\
	&-0.27   &\ 0.282 $\pm$ 0.066 $\pm$ 0.010 &\ 0.281 $\pm$ 0.056 $\pm$ 0.011& \\
	&\ 0.27  &\ 0.254 $\pm$ 0.066 $\pm$ 0.010 &\ 0.239 $\pm$ 0.056 $\pm$ 0.010&\ 0.067$\pm$ 0.120 $\pm$ 0.025 & -0.012 $\pm$ 0.101 $\pm$ 0.019 \\
	&\ 0.74  &\ 0.383 $\pm$ 0.059 $\pm$ 0.015 &\ 0.385 $\pm$ 0.051 $\pm$ 0.014& -0.096$\pm$ 0.107 $\pm$ 0.026 & -0.028 $\pm$ 0.092 $\pm$ 0.020 \\
	&\ 1.27  &\ 0.218 $\pm$ 0.185 $\pm$ 0.009 &\ 0.205 $\pm$ 0.148 $\pm$ 0.009& -0.133$\pm$ 0.331 $\pm$ 0.061 & -0.147 $\pm$ 0.260 $\pm$ 0.038 \\ \hline    \end{tabular}
    \caption{Longitudinal single- and double-spin asymmetries, $A_L$ and $A_{LL}$,  for $W^\pm$ production obtained from the STAR 2013 data sample, as well as the combination of 2013 with 2011+2012 results. The longitudinal single-spin asymmetry is measured for six decay positron or electron pseudorapidity intervals. The longitudinal double-spin asymmetry was determined in the same intervals and the results for the same absolute pseudorapidity value were combined. The systematic uncertainties include all contributions and thus also include the point-by-point correlated uncertainties from the relative luminosity and beam polarization measurements that are broken out separately in Figs.~\ref{fig:money} and~\ref{fig:moneyTh}. \label{tab:WAL2013} }
  \end{center}
\end{table*}

\begin{figure}[tb]
\centering
\includegraphics[width=8.6cm]{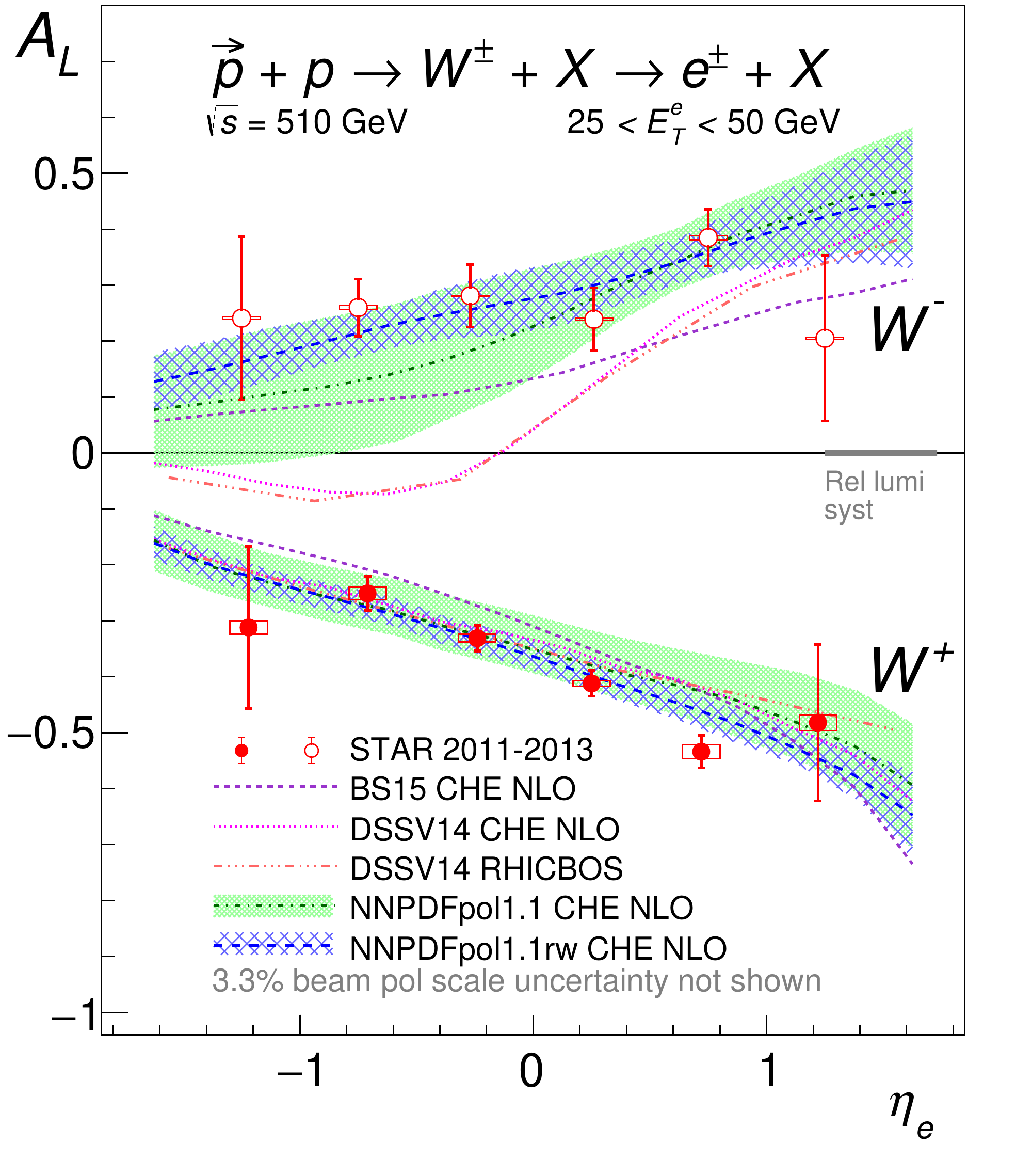}
\caption{\label{fig:moneyTh}Longitudinal single-spin asymmetries, $A_L$, for $W^\pm$ production as a function of the positron or electron pseudorapidity, $\eta_e$, for the combined STAR 2011+2012 and 2013 data samples for 25 $< E_T^e <$ 50\,GeV (points) in comparison to theory expectations (curves and bands) described in the text.}
\end{figure}

\begin{figure}[tb]
\centering
\includegraphics[width=8.6cm]{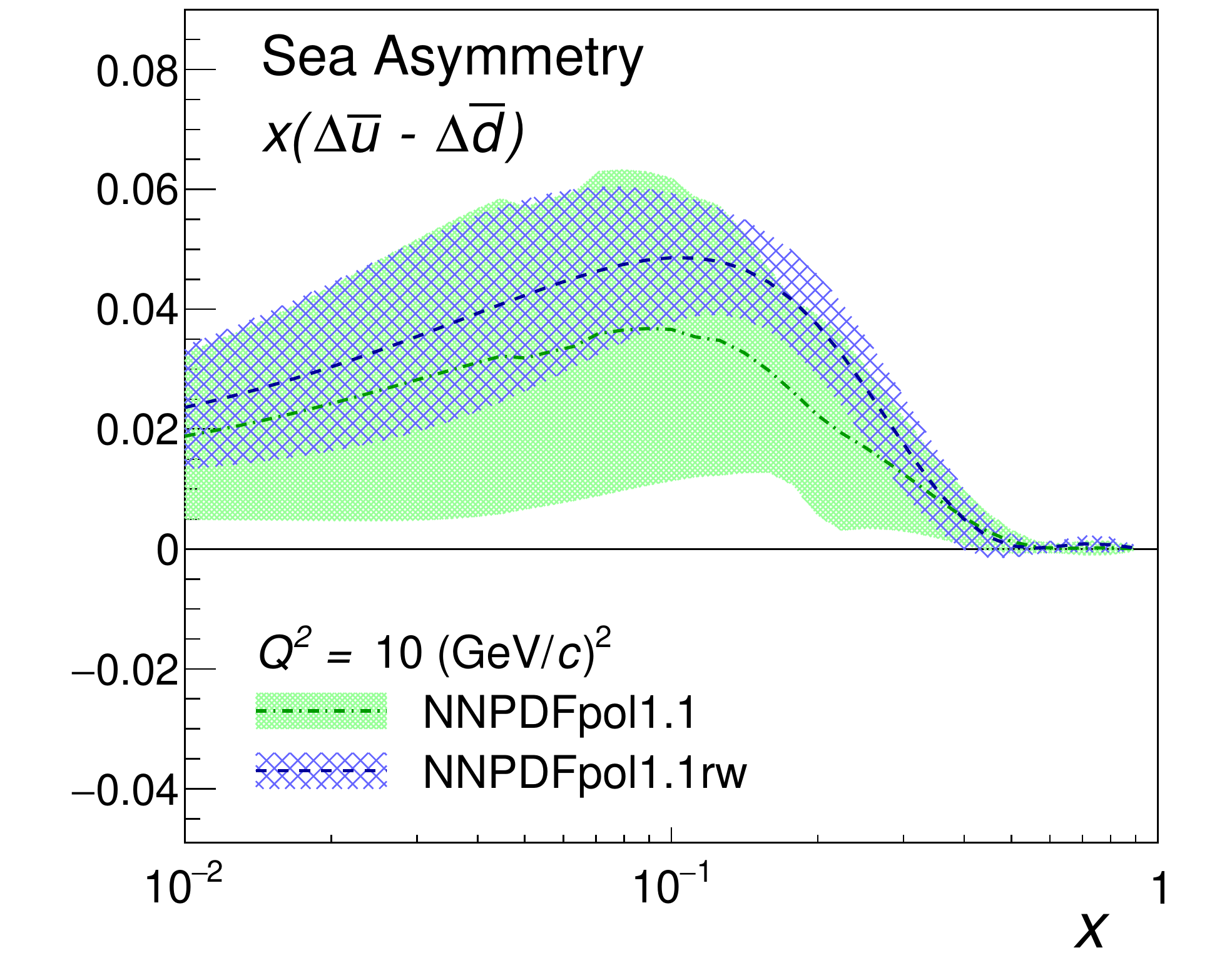}
\caption{\label{fig:PolSea}The difference of the light sea-quark polarizations as a function of $x$ at a scale of $Q^2$ = 10\,(GeV/$c$)$^2$.  The green band shows the NNPDFpol1.1 results~\cite{Nocera:2014gqa} and the blue hatched band shows the corresponding distribution after the STAR 2013 $W^\pm$ data are included by reweighting.}
\end{figure}

The new $W^\pm$ $A_L$ data are consistent with the previously published results, and have statistical uncertainties that are $40-50$\% smaller.
The combined STAR data are shown in Fig.~\ref{fig:moneyTh} and compared with expectations based on the DSSV14~\cite{deFlorian:2014yva}, NNPDFpol1.1~\cite{Nocera:2014gqa} and BS15~\cite{Bourrely:2015} PDFs evaluated using the next-to-leading order \textsc{CHE}~\cite{deFlorian:2010aa} and fully resummed \textsc{RHICBOS}~\cite{Nadolsky2003} codes.
The NNPDFpol1.1 analysis, unlike DSSV14 and BS15, includes the STAR 2011+2012 $W^\pm$ data~\cite{Adamczyk:2014xyw}, which reduces in particular the uncertainties for $W^-$ expectations at negative $\eta$.
To assess the impact, the STAR 2013 data were used in the reweighting procedure of Refs.~\cite{Ball:2010gb,Ball:2011gg} with the 100 publicly available NNPDFpol1.1 PDFs.
The results from this reweighting, taking into account the total uncertainties of the STAR 2013 data and their correlations~\cite{Sup:mat}, are shown in Fig.~\ref{fig:moneyTh} as the blue hatched bands.
The NNPDFpol1.1 uncertainties~\cite{Nocera:2014gqa} are shown as the green bands for comparison.
Figure~\ref{fig:PolSea} shows the corresponding differences of the light sea-quark polarizations versus $x$ at a scale of $Q^2 = 10\,(\mathrm{GeV}/c)^2$.
The data confirm the existence of a sizable, positive $\Delta\bar{u}$ in the range $0.05 < x < 0.25$~\cite{Adamczyk:2014xyw} and the existence of a flavor asymmetry in the polarized quark sea.

In addition, $ A_L$ was determined for $Z/\gamma^*$ production from a sample of 274 electron--positron pairs with 70 $<m_{e^+e^-}<$ 110\,GeV/$c^2$. The $e^+$ and $e^-$ were each required to be isolated, have $|\eta_e|<1.1$, and $E^e_T >$ 14\,GeV.
The result, $A^{Z/\gamma^*}_L = -0.04 \pm 0.07$, is consistent with that in Ref.~\cite{Adamczyk:2014xyw} but with half the statistical uncertainty. The systematic uncertainty is negligible compared to the statistical uncertainty.
This result is also consistent with theoretical expectations, $A^{Z/\gamma^*}_L = -0.08$ from DSSV14~\cite{deFlorian:2014yva} and $A^{Z/\gamma^*}_L =-0.04$ from  NNPDFpol1.1~\cite{Nocera:2014gqa}.

In summary, we report new STAR measurements of longitudinal single-spin and double-spin asymmetries for $W^\pm$ and single-spin asymmetry for $Z/\gamma^*$ bosons produced in polarized proton--proton collisions at $\sqrt{s}$ = 510\,GeV. The production of weak bosons in these collisions and their subsequent leptonic decay is a unique process to delineate the quark and antiquark polarizations in the 
proton by flavor. The $A_L$ data for $W^+$ and $W^-$, combined with previously published STAR results, show a significant preference for $\Delta\bar{u}(x,Q^2) > \Delta\bar{d}(x,Q^2)$ in the fractional momentum range 0.05 $< x <$ 0.25 at a scale of $Q^2 = 10$\,(GeV/$c$)$^2$. This is opposite to the flavor asymmetry observed in the spin-averaged quark-sea distributions. 

We thank the RHIC Operations Group and RCF at BNL, the NERSC Center at LBNL, and the Open Science Grid consortium for providing resources and support.  This work was supported in part by the Office of Nuclear Physics within the U.S. DOE Office of Science, the U.S. National Science Foundation, the Ministry of High Education and Science of the Russian Federation, National Natural Science Foundation of China, Chinese Academy of Science, the Ministry of Science and Technology of China and the Chinese Ministry of Education, the National Research Foundation of Korea, Czech Science Foundation and Ministry of Education, Youth and Sports of the Czech Republic, Department of Atomic Energy and Department of Science and Technology of the Government of India, the National Science Centre of Poland, the Ministry  of Science, Education and Sports of the Republic of Croatia, RosAtom of Russia and German Bundesministerium f\"ur Bildung, Wissenschaft, Forschung and Technologie (BMBF) and the Helmholtz Association.




\end{document}


Table~\ref{tab:CorrMtx} contains the correlations of the quadrature sum of the statistical and total systematic uncertainties on $A_L(\eta_e)$ for $W^-$ and $W^+$ from the STAR 2013 data.

\begin{table}[h]
    \centering
    \begin{tabular}{ c c | c c c c c c c c c c c c c} \hline\hline
    	&  & \multicolumn{6}{c}{ $W^-$ } & \multicolumn{6}{c}{ $W^+$ } \\ \hline
			& $\left<\eta_e\right>$ & -1.27 & -0.74 & -0.27 & 0.27 & 0.74 & 1.27 & -1.24 & -0.72 & -0.24 & 0.24 & 0.72 & 1.24 \\ \hline
  	\multirow{6}{*}{$W^-$} 
& -1.27& 1& 0.006& 0.009& 0.010& 0.019& 0.007& -0.229& -0.006& -0.021& -0.021& -0.021& -0.001\\
& -0.74& & 1& 0.024& 0.023& 0.041& 0.008& -0.013& \ 0.003& -0.022& -0.050& -0.027& -0.005\\
& -0.27& & & 1& 0.019& 0.037& 0.005& -0.014& -0.017& -0.041& -0.049& -0.055& -0.003\\
&\ 0.27& & & & 1& 0.031& 0.001& -0.011& -0.009& -0.029& -0.040& -0.039& -0.009\\
&\ 0.74& & & & & 1& 0.014& -0.017& -0.018& -0.054& -0.075& -0.062& -0.012\\
&\ 1.27& & & & & & 1& -0.002& -0.004& -0.008& -0.015& -0.014& -0.238\\
\multirow{6}{*}{$W^+$}
& -1.24& & & & & & & 1& \ 0.018& \ 0.035& \ 0.034& \ 0.040& \ 0.021\\
& -0.72& & & & & & & & 1& \ 0.150& \ 0.146& \ 0.209& \ 0.010\\
& -0.24& & & & & & & & & 1& \ 0.190& \ 0.232& \ 0.026\\
&\ 0.24& & & & & & & & & & 1& \ 0.244& \ 0.030\\
&\ 0.72& & & & & & & & & & & 1& \ 0.029\\
&\ 1.24& & & & & & & & & & & & 1\\ \hline
    \end{tabular}
    \caption{Bin-to-bin correlations of the quadrature sum of the statistical and systematic uncertainties on $A_L(\eta_e)$ for $W^-$ and $W^+$ from the STAR 2013 data. The values below the diagonal of this matrix are identical to those above the diagonal and are omitted for clarity.}
    \label{tab:CorrMtx}
\end{table}